# Absorption and fluorescence properties of oligothiophene biomarkers from long-range-corrected time-dependent density functional theory

*Bryan M. Wong*[a], Manuel Piacenza[b], and Fabio Della Sala[b]*

[a] Materials Chemistry Department, Sandia National Laboratories, Livermore, California 94551, USA

[b] National Nanotechnology Laboratory of CNR-INFM, Distretto Tecnologico ISUFI, Università del Salento, Via per Arnesano, I-73100 Lecce, Italy

*Corresponding author. E-mail: bmwong@sandia.gov.

The absorption and fluorescence properties in a class of oligothiophene push-pull biomarkers are investigated with a long-range-corrected (LC) density functional method. Using linear response time-dependent density functional theory (TDDFT), we calculate excitation energies, fluorescence energies, oscillator strengths, and excited-state dipole moments. To benchmark and assess the quality of the LC-TDDFT formalism, an extensive comparison is made between LC-BLYP excitation energies and approximate coupled cluster singles and doubles (CC2) calculations. When using a properly-optimized value of the range parameter, $\mu$, we find that the LC technique provides an accurate description of charge-transfer excitations as a function of biomarker size and chemical functionalization. In contrast, we find that re-optimizing the fraction of Hartree Fock exchange in conventional hybrid functionals still yields an inconsistent description of excitation energies and oscillator strengths for the two lowest excited states in our series of biomarkers. The results of the present study emphasize the importance of a distance-dependent contribution of exchange in TDDFT for investigating excited-state properties.

## 1. Introduction

Over the last few years, time-dependent density functional theory (TDDFT)[1-4] has made tremendous progress in the accurate description of electronic excitations and time-dependent quantum-mechanical phenomena. Based on the Runge-Gross theorem which relates time-dependent densities with time-dependent potentials,[5] TDDFT can, in principle, be applied to any time-dependent quantum-mechanical situation. For strong, time-dependent potentials, the full Kohn-Sham density must be obtained as a function of position and time. However, if the time-dependent potential is weak (i.e., in optical absorption), one can use linear response theory to obtain the excitation energies from the eigenvalues of a random-phase approximation (RPA)-like matrix equation.[1] As a result, linear-response TDDFT has become the method of choice for evaluating excited-state energies and properties of large molecular systems.

Despite the overwhelming success of TDDFT for predicting molecular excited states, it is well-known that an accurate description of long-range charge-transfer effects provides a significant challenge for the TDDFT formalism.[6-13] This shortcoming is not a failure of TDDFT itself (which is formally an exact theory), nor is it a breakdown of the linear-response approximation. As in the case for ground-state DFT, this limitation arises from approximations to the (still unknown) exact exchange-correlation functional. As shown by several groups, the use of conventional exchange-correlation functionals results in severely underestimated charge-transfer excitation energies and incorrect asymptotic potential energy surfaces resulting from electron-transfer self-interaction.[8,10,14,15] Although the use of hybrid functionals such as B3LYP[16] can partially reduce the self-interaction error, these conventional functionals still demonstrate a severe underestimation of excitation energies for long-range charge-transfer states.[17-20]

Recognizing the shortcomings of conventional functionals, major methodological progress has been made in DFT techniques which incorporate a position-dependent admixture of Hartree Fock (HF) exchange in the exchange-correlation functional.[11,21-35] Originally developed by Gill[21] and Savin,[22,23,27] these range-separated functionals partially account for long-range charge-separation effects by adding a growing fraction of exact exchange when the interelectronic distance increases (see section 2.2). This range-separation technique has been further modified and applied in many forms by Hirao *et al.*[11,24,28,31,32] in their LC (long-range-corrected) functional and by Handy *et al.*[26,30,35] with their CAM-B3LYP (Coulomb-attenuating method-B3LYP) methods. There has also been very recent work in constructing and providing diagnostic tests for new LC functionals. The Scuseria group has developed several new range-separated functionals based on a semilocal exchange approach.[33,36-38] Rohrdanz and Herbert have used this exchange-hole model to design a new functional which accurately describes both ground- and excited-states.[39] As a quantitative test of charge-transfer excitations, the Tozer group has rationalized the efficiency of the CAM-B3LYP method using a diagnostic test based on spatial overlap between orbitals involved in the excitation.[40] In terms of chemical applications, the Adamo and Scuseria groups have also presented benchmarks for several families of excitations including the electronic spectra of anthroquinone dyes,[41] $n \rightarrow \pi^*$ transitions in nitroso and thiocarbonyl dyes,[42] $\pi \rightarrow \pi^*$ excitations in organic chromophores,[43] and electronic transitions in substituted azobenzenes.[44]

In this work, we investigate the performance of various TDDFT methods on the excited-state properties for a series of functionalized oligothiophene biomarkers. In several previous investigations, we and our collaborators have shown that functionalized oligothiophene esters possess high fluorescence efficiencies, good optical stabilities, large Stokes shifts, and versatile color tunability in the entire visible range.[45-49] These features, along with their favorable binding to oligonucleotides and proteins, make oligothiophene-based biomarkers useful as fluorescent dyes in DNA and proteins.[48,49] As shown in Fig. 1, these oligothiophene derivatives can be chemically modified to create various push-pull systems. The thiolate ($SCH_3$) group attached in the $\alpha$ or $\beta$ position of the thiophene serves as the electron donor, and the *N*-succinimidyl ester or ethylamide group on the opposite side of the molecule corresponds to a change in the electron acceptor. As a result of this functionalization, these oligothiophene derivatives have already shown promising experimental results in fluorescence microscopy and protein-labeling.[48,49]

From a theoretical point of view, one would like to use a computational design beforehand to predict the effect of functionalization on the optical properties of these biomarkers before they are used as probes in fluorescent experiments. Indeed, in previous theoretical studies we have shown that wavefunction-based approximate coupled cluster singles and doubles (CC2) approaches were necessary to accurately predict the absorption and emission energies of charge-transfer transitions in these biomarkers.[45-47] Although there has been great progress in resolution-of-the-identity (RI) techniques for CC2 methods,[50] these wavefunction-based approaches are still computationally much more demanding than TDDFT methods. In this work, we test the performance of the LC-TDDFT approach for obtaining excited-state energies and properties for our oligothiophene markers. Our intent is not to resolve the many open questions regarding how best to formulate range-separated exchange-correlation functionals or providing universal benchmarks for all ideal chemical systems. The point of the calculations presented is to demonstrate that the LC-TDDFT formalism offers an efficient and accurate approach for describing optical properties in these new biomarker systems. In particular, these systems represent an excellent test set for investigating the accuracy of range-separated functionals in both delocalized and localized charge-transfer states. In all of these biomarkers, the $S_1$ excited state is dominated by a high-oscillator strength transition from the highest occupied molecular orbital (HOMO) to a *delocalized* lowest unoccupied molecular orbital (LUMO). In contrast, the $S_2$ excited state is a low-oscillator strength *charge-transfer* transition from the second highest occupied molecular orbital (HOMO-1), localized on the $SCH_3$ group, to the LUMO. Fig. 2 shows, as a specific example, the relevant frontier orbitals of NS-[2T]-$S_\alpha$ involved in the $S_1$ and $S_2$ excitations. It is, therefore, interesting to investigate whether a concurrent description of these two excited states of different character can be simultaneously described by the LC-TDDFT formalism. In the present study, we compute excitation energies, fluorescence energies, oscillator strengths, and excited-state dipole moments for each of the 12 oligothiophenes depicted in Fig. 1. Based on the overall trends in absorption and fluorescence properties, we find that the LC-TDDFT formalism significantly improves the description of excited-state properties in oligothiophene charge-transfer biomarkers compared to conventional hybrid functionals which incorporate a constant fraction of HF exchange.

## 2. Theory and Methodology

### 2.1 Conventional Hybrid Functionals

One of the most widely-used hybrid DFT schemes for the exchange-correlation energy is Becke's three-parameter B3LYP method[16] which is usually formulated as

$$E_{xc} = a_0 E_{x,\mathrm{HF}} + \left(1 - a_0\right) E_{x,\mathrm{Slater}} + a_x \Delta E_{x,\mathrm{Becke88}} + \left(1 - a_c\right) E_{c,\mathrm{VWN}} + a_c \Delta E_{c,\mathrm{LYP}}. \quad (1)$$

In this expression, $E_{x,\mathrm{HF}}$ is the HF exchange energy based on Kohn-Sham orbitals, $E_{x,\mathrm{Slater}}$ is the uniform electron gas exchange-correlation energy,[51] $\Delta E_{x,\mathrm{Becke88}}$ is Becke's 1998 generalized gradient approximation (GGA) for exchange,[52] $E_{c,\mathrm{VWN}}$ is the Vosko-Wilk-Nusair 1980 correlation functional,[53] and $\Delta E_{c,\mathrm{LYP}}$ is the Lee-Yang-Parr correlation functional.[54] The parameters $a_0 = 0.20$, $a_x = 0.72$, and $a_c = 0.81$ were determined by Becke using a least-squares fit to 56 experimental atomization energies, 42 ionization potentials, and 8 proton affinities.

Another common hybrid functional is Becke's half-and-half method (BHHLYP)[55] which makes use of Eq. (1) with $a_0 = 0.5$, $a_x = 0.5$, and $a_c = 0$. Depending on the choice of the GGA, there are numerous hybrid functionals in the literature which combine different GGA treatments of exchange and correlation with varying fractions, $a_0$, of HF exchange. To investigate the effect of modifying the HF exchange fraction on the optical properties of our biomarkers, we computed $S_1$ and $S_2$ vertical singlet excitation energies and fluorescence energies as a function of $a_0$ ranging from 0.0 to 1.0 in increments of 0.05. In these hybrid DFT benchmarks, we fixed $a_x = 1 - a_0$ in Eq. (1) but kept the correlation contribution with $a_c = 0.81$ unchanged. The $a_x = 1 - a_0$ convention is already a common choice used in many hybrid functionals[56-59] such as Becke's B1 convention[57] (in a separate study, we carried out calculations with $a_x$ fixed to the original 0.72 value recommended by Becke and found that the error in excitation energies was larger by 0.02 eV compared to the $a_x = 1 - a_0$ convention). It is also worth noting that the set of parameters $a_0 = 0.5$ and $a_x = 1 - a_0 = 0.5$ yields a functional similar to the BHHLYP functional ($a_c = 0$) with the exception that our choice has an extra correlation contribution due to the $\Delta E_{c,\mathrm{LYP}}$ term.

### 2.2 Long-Range Exchange Corrections

In contrast to conventional hybrid functionals which incorporate a constant fraction of HF exchange, the LC scheme for DFT[11,24,28,31,32] partitions the electron repulsion operator $1/r_{12}$ into short- and long-range components as

$$\frac{1}{r_{12}}=\frac{1-\mathrm{erf}\left(\mu r_{12}\right)}{r_{12}}+\frac{\mathrm{erf}\left(\mu r_{12}\right)}{r_{12}}. \tag{2}$$

The "erf" term denotes the standard error function, $r_{12}=\left|\mathbf{r}_1-\mathbf{r}_2\right|$ is the interelectronic distance between electrons at coordinates $\mathbf{r}_1$ and $\mathbf{r}_2$, and $\mu$ is an adjustable damping parameter having units of inverse Bohrs. The first term in Eq. (2) is a short-range interaction which decays rapidly on a length scale of ~ $2/\mu$, and the second term is the long-range "background" interaction.[24] For a pure density functional (i.e. BLYP or PBE) which does not already include a fraction of nonlocal HF exchange, the exchange-correlation energy according to the LC scheme is

$$E_{xc}=E_{c,\mathrm{DFT}}+E_{x,\mathrm{DFT}}^{\mathrm{SR}}+E_{x,\mathrm{HF}}^{\mathrm{LR}}, \tag{3}$$

where $E_{c,\mathrm{DFT}}$ is the DFT correlation functional, $E_{x,\mathrm{DFT}}^{\mathrm{SR}}$ is the short-range DFT exchange functional, and $E_{x,\mathrm{HF}}^{\mathrm{LR}}$ is the HF contribution to exchange computed with the long-range part of the Coulomb operator. The modified $E_{x,\mathrm{HF}}^{\mathrm{LR}}$ term can be analytically evaluated with Gaussian basis functions,[60] and the short-range $E_{x,\mathrm{DFT}}^{\mathrm{SR}}$ contribution is computed with a modified exchange kernel specific for each generalized gradient approximation (GGA). For the BLYP exchange-correlation functional used in this work, the short-range part of the exchange energy takes the form

$$E_{x,\mathrm{DFT}}^{\mathrm{SR}}=-\frac{1}{2}\sum_{\sigma}\int\rho_{\sigma}^{4/3}K_{\sigma}\left\{1-\frac{8}{3}a_{\sigma}\left[\sqrt{\pi}\,\mathrm{erf}\left(\frac{1}{2a_{\sigma}}\right)+2a_{\sigma}\left(b_{\sigma}-c_{\sigma}\right)\right]\right\}d^{3}r, \tag{4}$$

where $\rho_\sigma$ is the density of $\sigma$-spin electrons, and $K_\sigma$ is the GGA part of the exchange functional. The expressions for $a_\sigma$, $b_\sigma$, and $c_\sigma$ are given by

$$a_{\sigma}=\frac{\mu K_{\sigma}^{1/2}}{6\sqrt{\pi}\rho_{\sigma}^{1/3}}, \tag{5}$$

$$b_\sigma = \exp\left(-\frac{1}{4a_\sigma^2}\right) - 1, \tag{6}$$

and

$$c_\sigma = 2a_\sigma^2 b_\sigma + \frac{1}{2}. \tag{7}$$

The correlation contribution represented by $E_{c,\mathrm{DFT}}$ in Eq. (3) is left unmodified from its original DFT definition.

The key improvement in the LC scheme is the smooth separation of DFT and nonlocal HF exchange interactions through the parameter $\mu$. Specifically, the exchange-correlation potentials of conventional density functionals exhibit the wrong asymptotic behavior, but the LC scheme ensures that the exchange potential smoothly recovers the exact $-1/r$ dependence at large interelectronic distances. For extended charge-transfer processes, the long-range exchange corrections become particularly vital since these excitations are especially sensitive to the asymptotic part of the nonlocal exchange-correlation potential. In the conventional LC scheme used here, the damping parameter $\mu$ determines the relative contributions of DFT and HF to the exchange-correlation energy. For $\mu = 0$, Eq. (3) reduces to $E_{xc} = E_{c,\mathrm{DFT}} + E_{x,\mathrm{DFT}}$, and all electronic interactions are described with a pure exchange-correlation density functional. Conversely, the $\mu \to \infty$ limit corresponds to an exchange-correlation functional of the form $E_{xc} = E_{c,\mathrm{DFT}} + E_{x,\mathrm{HF}}$ where all DFT exchange has been replaced by nonlocal HF exchange.

To explore the effect of range-separated exchange on the optical properties of our biomarkers, we computed $S_1$ and $S_2$ vertical singlet excitation energies and fluorescence energies as a function of $\mu$ ranging from 0 to 0.90 Bohr$^{-1}$ (in increments of 0.05 Bohr$^{-1}$) while keeping the correlation contribution $E_{c,\mathrm{DFT}}$ unchanged. The result of varying the exchange contribution in the LC scheme is more general than conventional hybrid functionals which are defined with a fixed fraction of nonlocal HF exchange (i.e. B3LYP or PBE0). That is, conventional hybrids incorporate a constant admixture of HF exchange while the LC formalism mixes exchange energy densities based on interelectronic distances at each point in space.

### 2.3 Computational Details

For the oligothiophene biomarkers in this work, we benchmarked the performance of a long-range-corrected LC-BLYP functional against B3LYP, BHHLYP, and existing wavefunction-based CC2 calculations. To investigate the role of different HF exchange schemes in the LC and hybrid functionals (discussed further in section 3), we also explored the effect of varying the range parameter $\mu$ in LC-BLYP and the result of changing the HF exchange fraction, $a_0$, in the B3LYP functional. In order to maintain a consistent comparison across the LC-BLYP, B3LYP, BHHLYP, and CC2 levels of theory, unmodified geometries obtained from a previous work[47] were used for each of the four methods. The ground- and excited-state geometries from the previous study were optimized at the B3LYP/TZVP level of theory using DFT and TDDFT, respectively, and further details of the structures can be found in Ref. 47. In our TDDFT calculations, the two lowest singlet vertical excitations were calculated using an augmented TZVP basis set, further denoted as ATZVP. The custom ATZVP basis set employs standard TZVP basis functions with additional diffuse functions for second- and third-row atoms. The exponents for the ATZVP diffuse functions were obtained from a geometric series based on the original TZVP set and can be found in Ref. 47. All fluorescence electronic transitions were calculated as vertical de-excitations based on the TDDFT B3LYP/TZVP-optimized geometries of the $S_1$ state.

For the CC2 benchmark properties also obtained from Ref. 47, the resolution of the identity approximation (RI-CC2)[50] was used in conjunction with the ATZVP basis to calculate vertical excitations on the B3LYP/TZVP-optimized geometries. Throughout this work we use the CC2 energies as reference values for assessing the quality of the various TDDFT methods. We have previously found that the CC2 method correctly reproduces CASPT2 emission energies for unsubstituted bithiophene[61] and terthiophene[62] within 0.1 eV and 0.2 eV, respectively.[63] For absorption energies, a complete CASPT2 study of these systems is not available since small changes in the ground-state torsional angles can yield large variations (0.2 – 0.3 eV[61,62]) in the excitation energies. Moreover, Ref. 47 demonstrates that the CC2 method reproduces experimental differences between oligothiophene systems within 0.1 eV, despite a systematic overestimation of absolute excitation energies mainly related to the neglect of vibrational and solvent effects.[45] We therefore take the CC2 results as reliable reference values considering also that CC3 and CASPT2 calculations are out of reach for our large non-symmetric systems. As an additional check on the quality of the CC2 calculations, we found that none of the oligothiophene systems required a multi-reference treatment of electron correlation (D1 diagnostic values were in the 0.08 – 0.10 range), and contributions from single excitations were always greater than 90%.

For both the ground-state and TDDFT single-point calculations, we used a high-accuracy Lebedev grid consisting of 96 radial and 302 angular quadrature points. The ground- and excited-state LC-BLYP electric dipole moments were evaluated using analytical LC-TDDFT energy derivatives recently implemented by Chiba *et al.*[32] All *ab initio* calculations were performed with a locally modified version of GAMESS.[64]

## 3. Results

### 3.1 Excitation Energies

Figs. 3(a) and 3(b) display, as a function of $\mu$ and $a_0$, the $S_1$ and $S_2$ excitation energies of the BC-[3T]-$S_\beta$ biomarker compared against the CC2 calculations of Ref. 47. The corresponding figures for the other 11 molecules are very similar and can be found in the Electronic Supplementary Information (Figs. ESI-1 and ESI-2). The horizontal lines represent the CC2/ATZVP excitation energies, and the curved lines denote the TDDFT/ATZVP calculations. The most important features of these results show that *both* the LC-BLYP $S_1$ and $S_2$ excitation energies coincide with their respective CC2 reference values within a very small $\mu$-range of 0.28 $Bohr^{-1} < \mu < 0.37$ $Bohr^{-1}$. In stark contrast, Figs. 3(b) and ESI-2 show that there is not a single value or small range of $a_0$ in the B3LYP-like functional which gives reasonable accuracy for both $S_1$ and $S_2$ energies. The general trend for the B3LYP-like calculations is that the optimal value of $a_0$ for $S_1$ energies ($a_0 \sim 0.8$) is considerably larger than the optimal value for $S_2$ energies ($a_0 \sim 0.4$). In particular, the excitation energies obtained at a value of $a_0$ which is optimal for $S_1$ would give large errors in $S_2$ excitation energies, and vice versa.

### 3.2 Fluorescence Energies

Next, we consider fluorescence energies calculated from energy differences between the optimized $S_1$ state and the $S_0$ ground state (at the same reference $S_1$ geometry). Unlike the ground-state geometries which have inter-thiophene torsional angles between 16 and 24°, the optimized $S_1$ geometries are significantly more planar with inter-thiophene dihedral angles less than 6° (see Ref. 47). This effect arises from an increased quinoid character (antibonding interactions in the thiophene C=C bonds and bonding interactions in C–C bonds connecting thiophene rings) in the excited $S_1$ state. As a result, there is an enhanced rigidity of the molecular backbone accompanied by a substantial shortening of the inter-ring bond lengths upon electronic excitation.

Figs. 4(a) and 4(b) display, as a function of $\mu$ and $a_0$, the $S_1$ fluorescence energy of BC-[3T]-$S_\beta$ against the CC2 calculation of Ref. 47; fluorescence energy curves as a function of $\mu$ and $a_0$ for the other 11 biomarkers are available in the Electronic Supplementary Information. Compared to the $S_1$/$S_2$ excitation energies, the LC-BLYP fluorescence curves exhibit a weaker dependence on the range-separation parameter $\mu$ with an overall variation of ~0.6 eV. Surprisingly, Figs. 4(b) and ESI-4 show that the B3LYP-like functional generally yields large errors in fluorescence energies irrespective of the percentage of HF exchange included in the hybrid functional.

### 3.3 Optimal Values of $\mu$ and $a_0$

Using the CC2 excitation and fluorescence energies as reference values, we performed a total root-mean-square error (RMSE) analysis for all 36 energies (12 $S_1 \rightarrow S_0$, 12 $S_1 \leftarrow S_0$ and 12 $S_2 \leftarrow S_0$ transitions) as a function of $\mu$ and $a_0$. As seen in Fig. 5(a), the RMSE curve for LC-BLYP has a minimum at $\mu = 0.31$ Bohr$^{-1}$ with an RMS error of 0.12 eV. Perhaps surprisingly, this RMSE-optimized value of $\mu$ is quite close to the 0.33 Bohr$^{-1}$ value recommended by Iikura *et al.*[24] for ground-state properties. The RMSE curve in Fig. 5(b) for the B3LYP-like functional has a minimum at $a_0$ = 0.49, with a slightly larger error of 0.21 eV. We denote this re-optimized hybrid functional with $a_0$ = 0.49, $a_x$ = 1 – $a_0$ = 0.51, and $a_c$ = 0.81 as B3LYP* in the remainder of this work. Unless otherwise noted, all further LC-BLYP calculations indicate a range-separation parameter of $\mu$ set to 0.31 Bohr$^{-1}$.

Table 1 compares excited-state energies and oscillator strengths between B3LYP, B3LYP*, LC-BLYP, and CC2 for the bithiophene systems, and Table 2 gives the corresponding results for the terthiophene systems. Since the BHHLYP functional gives nearly identical results to the B3LYP* calculations, all BHHLYP values are listed in Tables ESI-1, ESI-2, and ESI-3 in the Electronic Supplementary Information. Figs. 6 and 7 depict in more detail the general trend in the $S_1 \leftarrow S_0$ and $S_2 \leftarrow S_0$ transition energies between the various TDDFT and CC2 results. The diagonal line in all of these figures represents an ideal 100% agreement between the CC2 energies and the corresponding TDDFT results. In Figs. 6(b) and 7(b), we also plot the excitation energies obtained with $\mu$ = 0.47 Bohr$^{-1}$ which is a recent re-parameterization used by Song *et al.*[65] for reaction barrier heights (the LC-BLYP$_{\mu=0.47}$ energies are also listed in Tables ESI-1, ESI-2, and ESI-3 in the Electronic Supplementary Information).

Figs. 6 and 7 show that the LC-BLYP$_{\mu=0.31}$ calculations are in excellent agreement with the CC2 results for both $S_1$ and $S_2$ excitations, while the B3LYP functional severely underestimates excitation energies for all 12 of the oligothiophenes. This systematic underestimation of $S_1$ excitation energies is significantly improved upon using the B3LYP*/BHHLYP functionals; however, it is apparent from Fig. 7(a) that several of the $S_2 \leftarrow S_0$ transition energies are severely *overestimated* at the B3LYP* and BHHLYP levels of theory. As a result, both the B3LYP* and BHHLYP $S_2$ excitations increase too rapidly as a function of energy (a least-squares fit yields a slope of 1.4) in comparison to the CC2 benchmark results (slope ≡ 1). The LC-BLYP$_{\mu=0.47}$ calculations also considerably overestimate the $S_2 \leftarrow S_0$ transition energies. More interestingly, using a simple linear fit to the $S_1$ data points, one obtains high statistical correlations ($R^2$ = 0.97 – 1.00) for all functionals, indicating a simple, systematic error in these excitation energies. In contrast, the same linear fitting procedure for the $S_2$ excitations only yields high correlations for LC-BLYP$_{\mu=0.31}$ ($R^2$ = 0.96) with tremendously poorer $R^2$ values for B3LYP*/BHHLYP ($R^2$ = 0.78), and B3LYP ($R^2$ = 0.79). Among the oligothiophene biomarkers studied here, the overall accuracy in excitation energies is greatly improved with the LC scheme while the hybrid functionals are unable to reproduce general trends in $S_2$ excitations *even if the fraction, $a_0$, of HF exchange is optimized.*

Table 3, and Fig. 8 compare fluorescence energies and properties between B3LYP, B3LYP*, LC-BLYP, and CC2 for all 12 oligothiophene systems. As expected from our previous analysis of excitation energies, the fluorescence energies in Fig. 8(a) are significantly underestimated by the B3LYP calculations. However, as found for the $S_1$ excitation energies, all TDDFT methods show a high degree of statistical correlation with the CC2 reference values ($R^2$ = 0.95 – 1.00). In recent studies, the Adamo group also found that the LC scheme provided consistent $R^2$ values (compared to traditional hybrid functionals) of excitation energies in conjugated systems.[41,42] Our benchmarks support their suggestion and also show that the LC treatment provides a more consistent picture for fluorescence energies.

**4. Discussion**

On a qualitative level, all of the theoretical methods reproduce the expected trend that the excitation energies of the terthiophene derivatives become reduced relative to the corresponding bithiophene biomarkers. Introducing the electron-donating $SCH_3$ substituent in either the $\alpha$ or $\beta$ position decreases the energy gap between the HOMO and LUMO and, therefore, also reduces the absorption and fluorescence energies. Alternatively, changing the ethylamide group with the more electronegative *N*-succinimidyl ester corresponds to a change of the electron acceptor in these biomarkers. As a result of this functionalization, the absorption energies of the ethylamide thiophenes become blue-shifted by 0.1–0.2 eV compared to their corresponding *N*-succinimidyl counterparts. This same trend can also be seen in the fluorescence data, although the deviations between the ethylamide and *N*-succinimidyl thiophene energies are significantly smaller.

Despite the common prediction of chemical-functionalization trends, the performance of the various TDDFT methods relative to the CC2 calculations is considerably different, particularly for the charge-transfer transitions. To explain these trends and to put our discussion of charge-transfer effects on a more quantitative basis, we also examined the ground- and excited-state dipoles for all 12 oligothiophenes. The difference in dipole moment between the ground and excited state directly reflects the extent of charge transfer involved in the absorption/fluorescence process. Table 4 compares ground- and excited-state dipoles between B3LYP, B3LYP*, and LC-BLYP for the $S_1/S_2 \leftarrow S_0$ absorption process, and Table 5 lists the corresponding results for the $S_1 \rightarrow S_0$ emission calculations. Again, the BHHLYP functional gives nearly identical results to the B3LYP* calculations, so we only discuss the B3LYP* results here. In comparing the results of Tables 4 and 5, we observe two general trends: (i) the B3LYP excited-state dipoles are significantly overestimated relative to the other TDDFT methods, and (ii) the B3LYP* dipoles for the $S_1$ excitations agree well with the LC-BLYP results, but there are significant deviations between the B3LYP* and LC-BLYP $S_2$ dipoles. We discuss both of these trends in detail below.

Of all the TDDFT methods examined in this study, the B3LYP functional exhibits the largest variation in excitation energies and dipoles. The B3LYP functional incorporates a fixed fraction of 20% HF exchange and, therefore, exhibits a $-0.2/r$ dependence for the exchange potential at large interelectronic distances. As a result, this incorrect exchange potential is not attractive enough, leading to an overestimation of electron transfer and hence a larger dipole moment. In particular, the B3LYP functional predicts an unphysical large dipole moment of 22.31 Debye for the $S_2$ excited state of NS-[3T]-$S_\alpha$, while the LC-BLYP functional predicts a significantly smaller dipole of 3.12 Debye. The qualitative description of charge-transfer excitations predicted by the B3LYP functional is especially inconsistent within the bithiophene systems. Specifically, we draw attention to the excitation energies and oscillator strengths of NS-[2T]-$S_\beta$ and BC-[2T]-$S_\beta$ reported in Table 1. At the CC2 level of theory, the $S_1$ excitations for both of these systems have large oscillator strengths and are characterized by single-particle transitions from the HOMO to the LUMO. The CC2 $S_2$ excitations, in contrast, have small oscillator strengths and are largely described by transitions from the HOMO-1 to the LUMO (cf. Fig. 2). At the B3LYP level of theory, however, the trend is reversed with the $S_2$ state having the larger oscillator strength for both NS-[2T]-$S_\beta$ and BC-[2T]-$S_\beta$ since the HOMO $\rightarrow$ LUMO and HOMO-1 $\rightarrow$

LUMO transitions are significantly mixed in both of these systems.[47] As a result of this mixing, the charge-transfer character of the $S_2$ state is diminished, and the B3LYP $S_2$ dipole moments listed in Table 1, are actually underestimated. The B3LYP* and LC-BLYP results, in contrast, do not exhibit these inconsistencies, and both the excitation energies and oscillator strengths predicted by B3LYP* and LC-BLYP are in exceptional agreement with the CC2 results.

The second discussion point concerns the evaluation of excited-state properties with the B3LYP* functional which incorporates a fixed fraction of 49% HF exchange. The larger percentage of exchange in B3LYP* widens the HOMO-LUMO gap which correspondingly increases the $S_1$ excitation energies towards the CC2 benchmark values. However, as seen in Fig. 7(a) and Table 1, the B3LYP* $S_2$ charge-transfer excitation energies become severely overestimated for the bithiophene systems. The large B3LYP* $S_2$ dipole moments listed in Table 4 also reflect this trend, particularly for the NS-[2T]-$S_\alpha$ and BC-[2T]-$S_\alpha$ systems. As a result, it is apparent that a delicate balance between exchange and correlation errors is necessary to simultaneously describe both the $S_1$ and $S_2$ excitations in these systems with reasonable accuracy. Although an increased percentage of HF exchange improves the $S_1$ excitation energies, this modification in B3LYP* overcompensates for the error in calculating $S_2$ properties. In particular, it is not possible to simultaneously obtain *both* accurate excitation energies and reasonable $R^2$ values by adjusting the fraction of HF exchange in B3LYP. As a result, we find that a distance-dependent contribution of HF exchange is required to accurately describe both the $S_1$ and $S_2$ excitations in these oligothiophene biomarkers.

## 5. Conclusion

In this study, we have extensively investigated the absorption and fluorescence properties in a series of functionalized oligothiophene derivatives which can be used as fluorescent biomarkers. For each of the 12 oligothiophenes, excited-state energies and properties were obtained using the linear-response formalism of TDDFT in conjunction with a functional modified specifically for long-range charge-transfer. To investigate the optimal value of the range-separation parameter, $\mu$, an extensive comparison was made between LC-BLYP excitation energies and CC2 calculations. Using this optimized value of $\mu$, we find that the range-separated LC-BLYP functional significantly improves the poor description given by hybrid functionals and provides a more consistent picture of excitation energies as a function of molecular size and structural modification.

Among the oligothiophene biomarkers studied here, we also calculated a large increase in the $S_2$ electric dipole moment with respect to that of the ground state, indicating a sizable charge transfer associated with the $S_2 \leftarrow S_0$ excitation. The amount of charge transfer involved in this electronic transition is significantly overestimated by B3LYP, leading to large dipole moments and inconsistent oscillator strengths for the bithiophene systems. Re-optimizing the percentage of HF exchange in B3LYP* does improve the description of some $S_1$ properties; however, the same procedure also corrupts the balance between exchange and correlation errors with several of the $S_2$ excitation energies becoming severely overestimated. In particular, we find that conventional hybrid functionals are unable to reproduced general trends in both $S_1$ and $S_2$ excitations even if the fraction of HF exchange is optimized. The LC-BLYP results, in contrast, do not exhibit these inconsistencies, and the excitation energies and trends predicted by LC-BLYP are in exceptional agreement with the CC2 results.

In closing, the present study clearly indicates that long-range exchange corrections play a vital role in predicting the excited-state dynamics of oligothiophene biomarker systems. In contrast to conventional hybrids like B3LYP and BHHLYP which incorporate a constant percentage of HF exchange, we find that a distance-dependent contribution of HF exchange is required to simultaneously describe both the $S_1$ and $S_2$ excitations in these oligothiophenes. We are currently investigating how chemical binding to other biomolecules might change the optical properties of our oligothiophene biomarkers. With this in mind, we anticipate that the LC-TDDFT technique will play a significant role in predicting the different photophysical properties of these systems.

**Acknowledgements**

Sandia is a multiprogram laboratory operated by Sandia Corporation, a Lockheed Martin Company, for the United States Department of Energy's National Nuclear Security Administration under contract DE-AC04-94AL85000. This work is also partially funded by the ERC-Starting Grant FP7-Project "DEDOM," grant agreement number 207441.

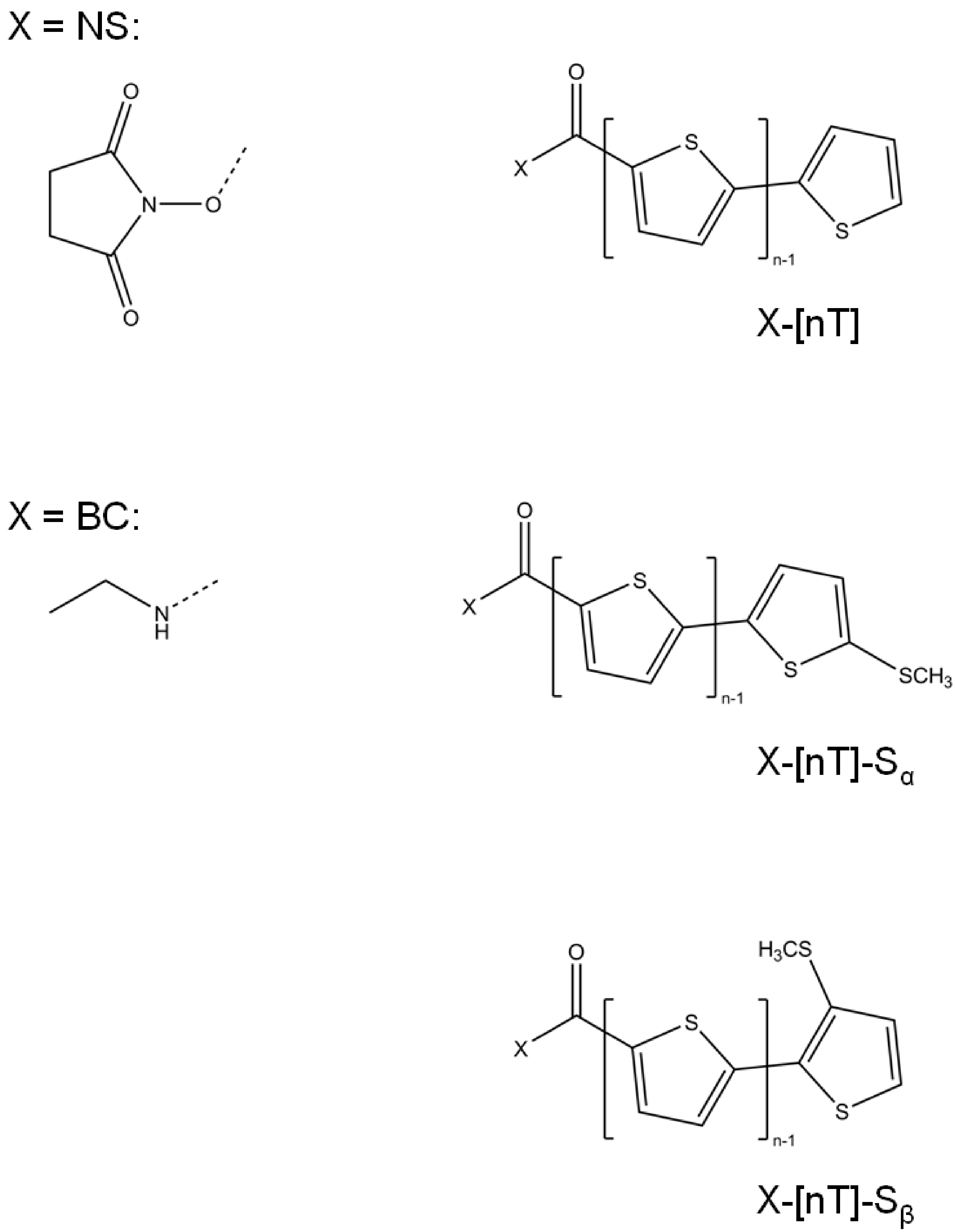

**Fig. 1** Molecular structures of the bithiophene ($n$ = 2) and terthiophene ($n$ = 3) systems. The *N*-succinimidyl esters are labeled with the prefix NS, and the ethylamide systems are denoted by the prefix BC. The thiophene systems which have been functionalized with thiolate groups are labeled with either the $S_\alpha$ or $S_\beta$ suffixes.

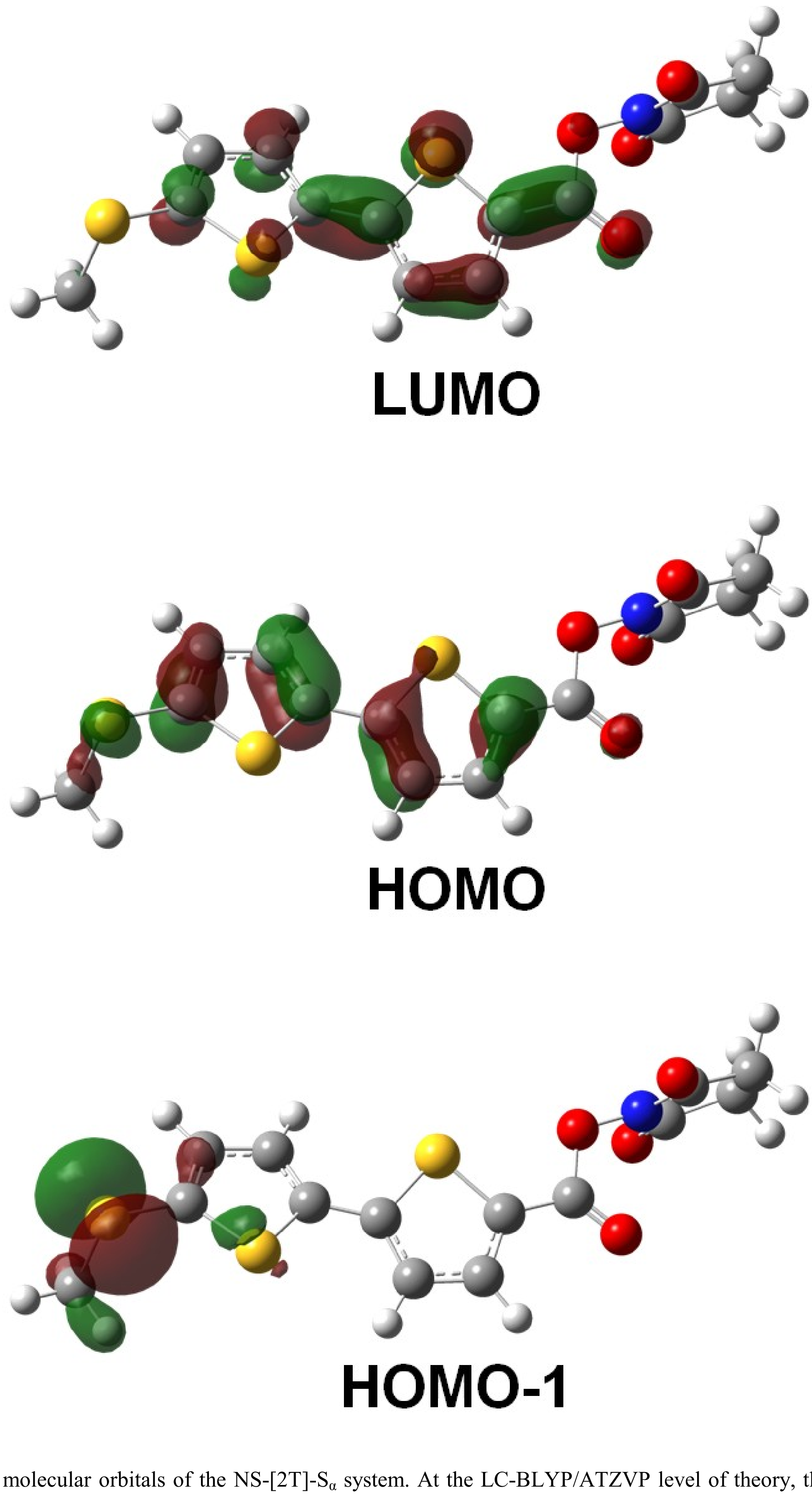

**Fig. 2** Frontier molecular orbitals of the NS-[2T]-$S_\alpha$ system. At the LC-BLYP/ATZVP level of theory, the $S_1$ excitation is primarily characterized by a HOMO → LUMO single-particle transition. In contrast, the $S_2$ transition is largely described by a HOMO-1 → LUMO charge-transfer excitation.

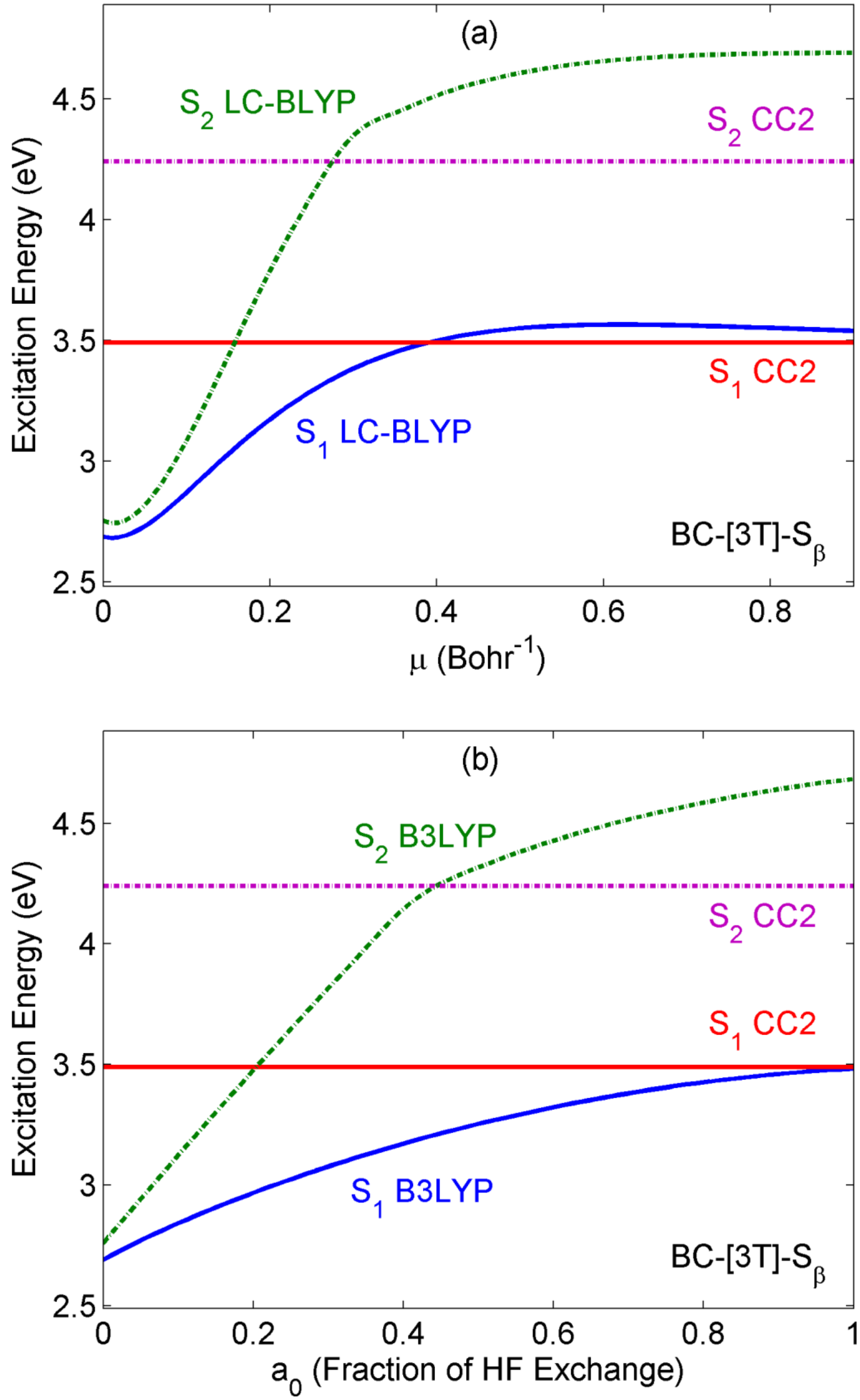

**Fig. 3** $S_1 \leftarrow S_0$ and $S_2 \leftarrow S_0$ vertical excitation energies for the BC-[3T]-$S_\beta$ biomarker as a function of (a) the LC-BLYP range parameter $\mu$ and (b) the HF exchange fraction $a_0$ in a B3LYP-like hybrid functional. The horizontal lines represent the CC2/ATZVP excitation energies, and the curved lines denote the TDDFT/ATZVP calculations. The solid lines denote $S_1 \leftarrow S_0$ excitation energies while dashed lines represent $S_2 \leftarrow S_0$ excitations.

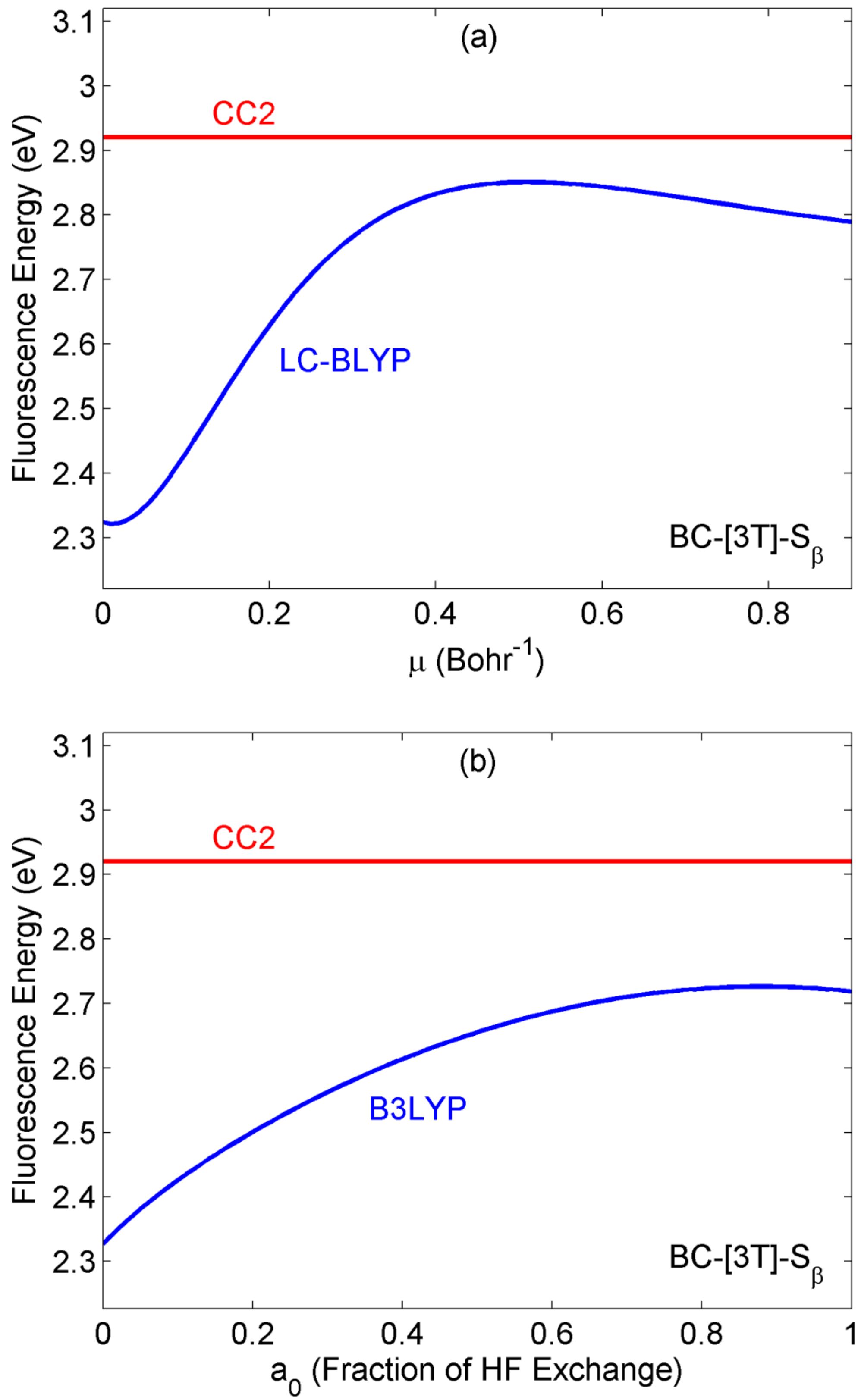

**Fig. 4** $S_1 \rightarrow S_0$ fluorescence energies for the BC-[3T]-$S_\beta$ biomarker as a function of (a) the LC-BLYP range parameter $\mu$ and (b) the HF exchange fraction $a_0$ in a B3LYP-like hybrid functional. The horizontal line represents the CC2/ATZVP excitation energy, and the curved line denotes the TDDFT/ATZVP calculations.

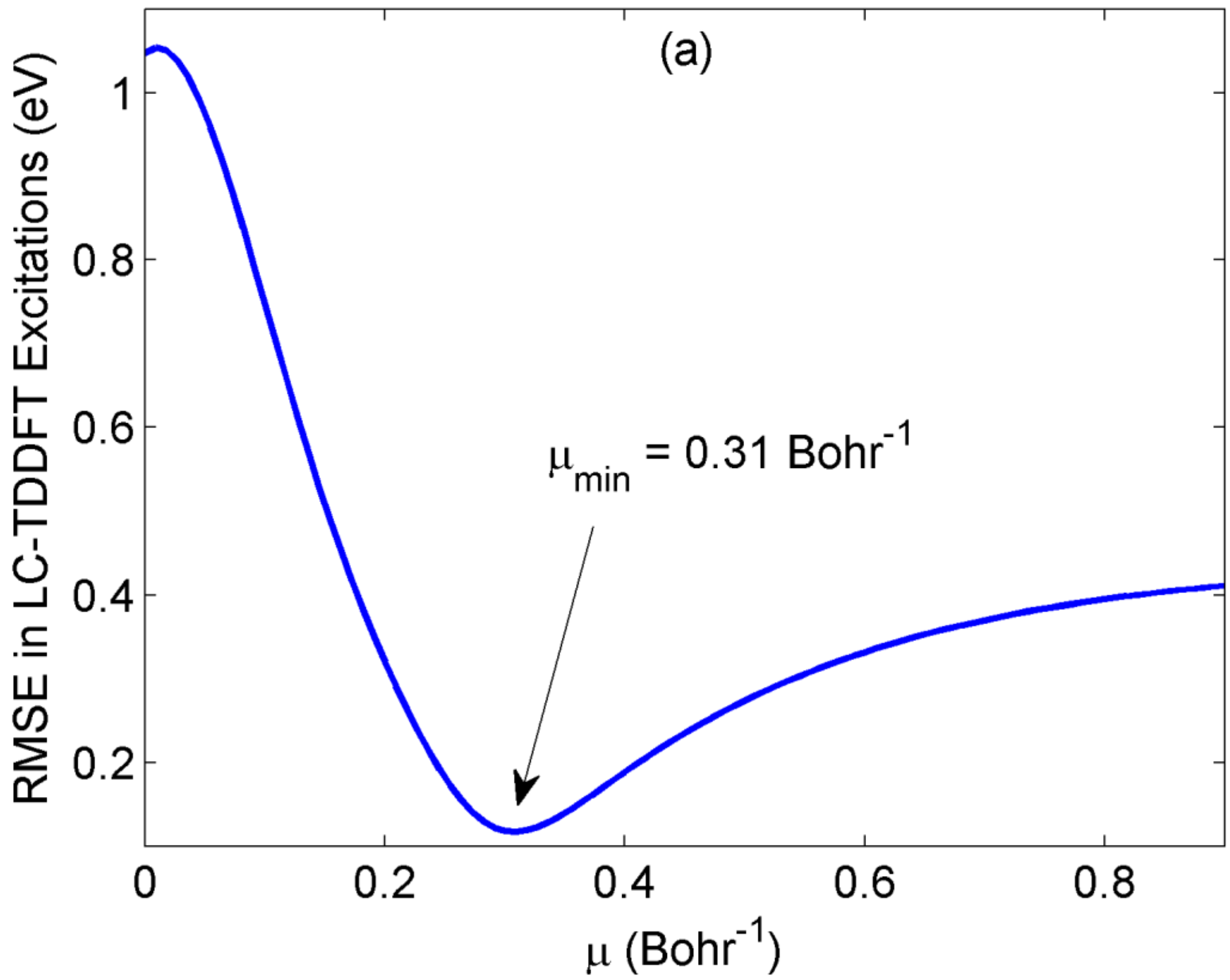

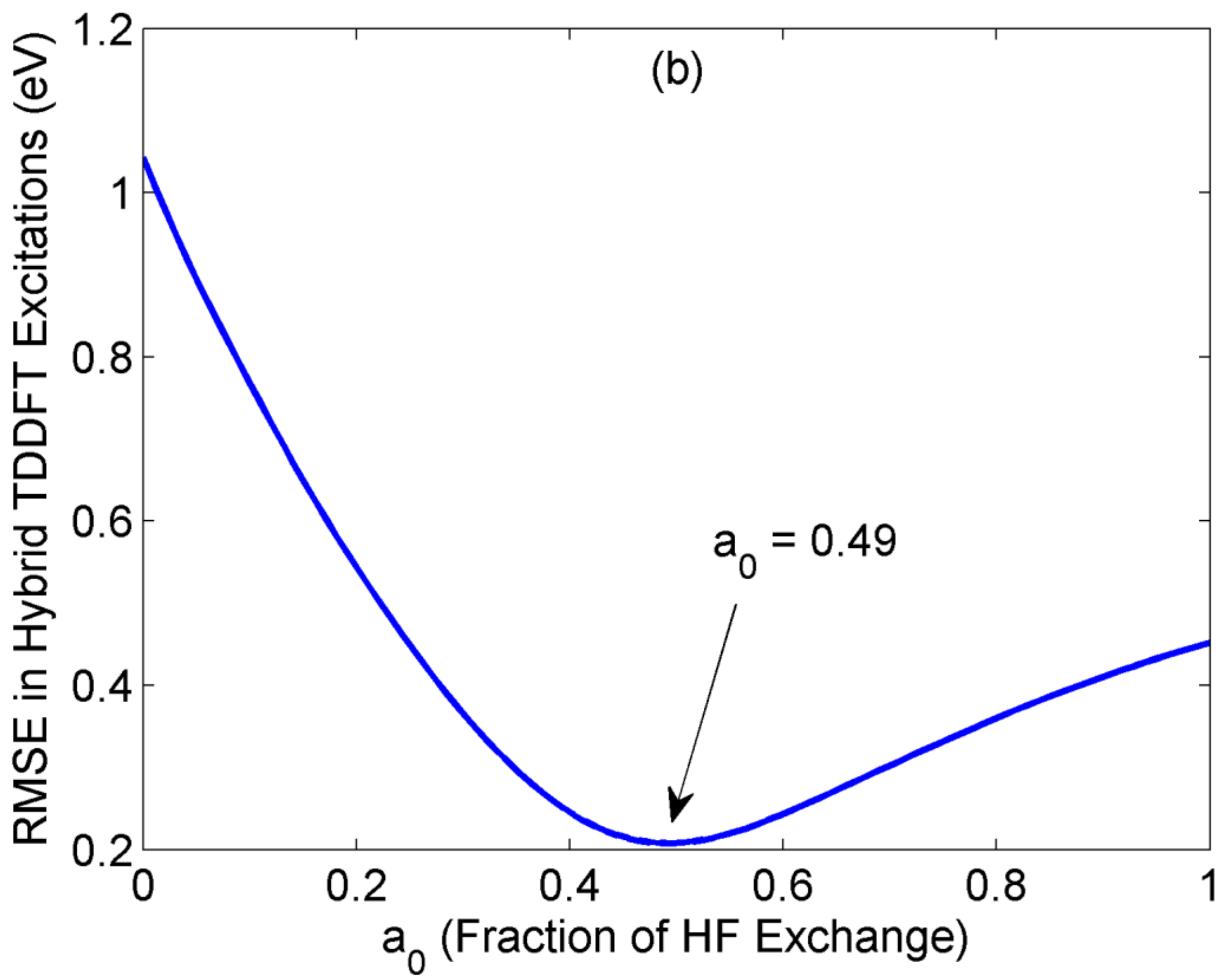

**Fig. 5** Total root-mean-square errors (RMSE) as a function of (a) the range parameter $\mu$ in the LC-BLYP functional and (b) the HF exchange fraction $a_0$ in a B3LYP-like hybrid functional. Fig. 3(a) shows the RMSE curve having a minimum at $\mu$ = 0.31 Bohr$^{-1}$, and Fig. 3(b) shows the RMSE curve having a minimum at $a_0$ = 0.49.

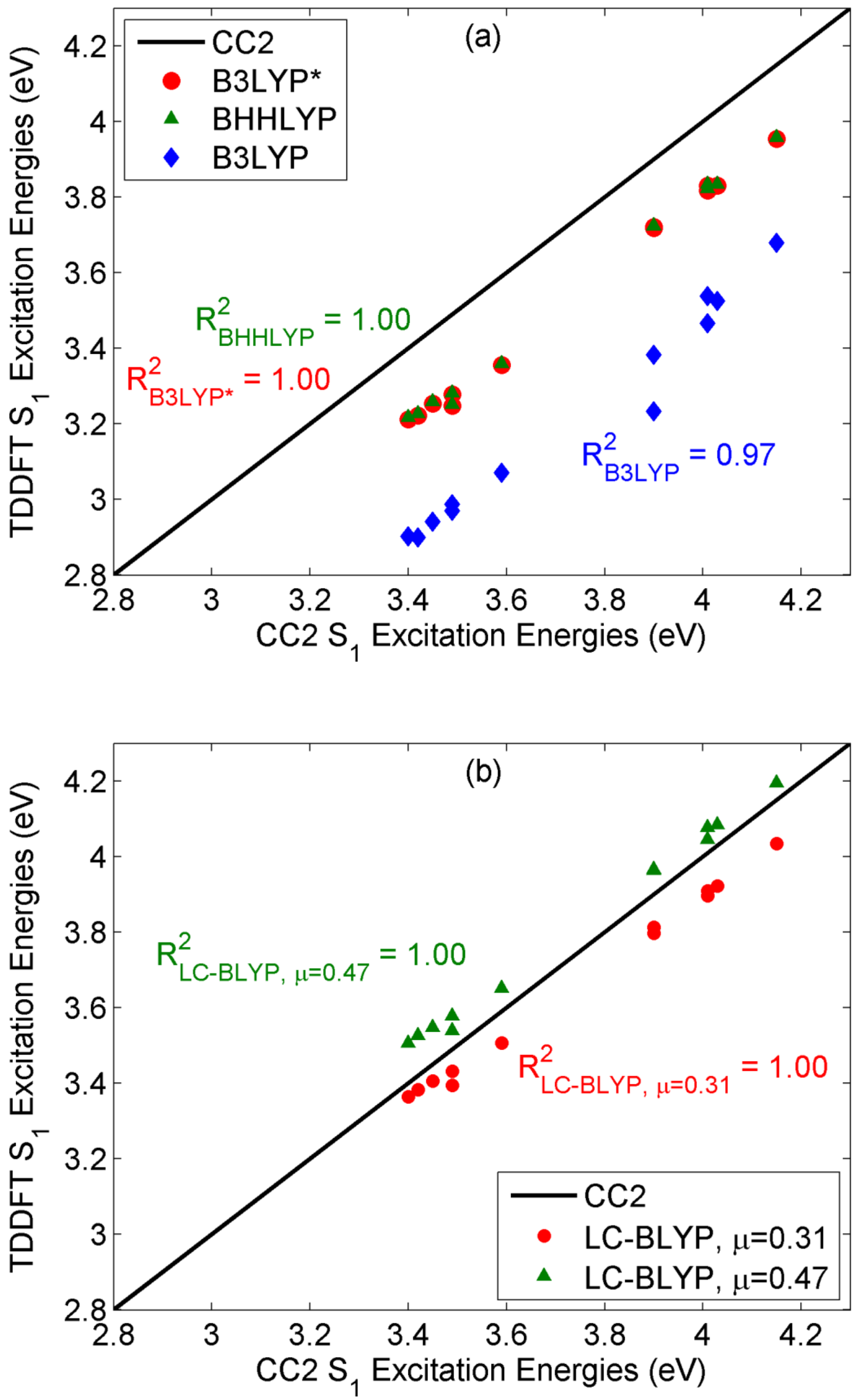

**Fig. 6** Comparison between TDDFT and CC2 $S_1 \leftarrow S_0$ excitation energies for (a) conventional hybrid functionals and (b) range-separated LC-BLYP functionals. The diagonal line in each figure represents a perfect match between CC2 and TDDFT $S_1 \leftarrow S_0$ excitation energies. The $R^2$ values were obtained from a simple linear fit to the data points themselves and not calculated with respect to the diagonal lines shown in the figures. In Fig. 6(a), the B3LYP* functional gives nearly identical results to the BHHLYP calculations.

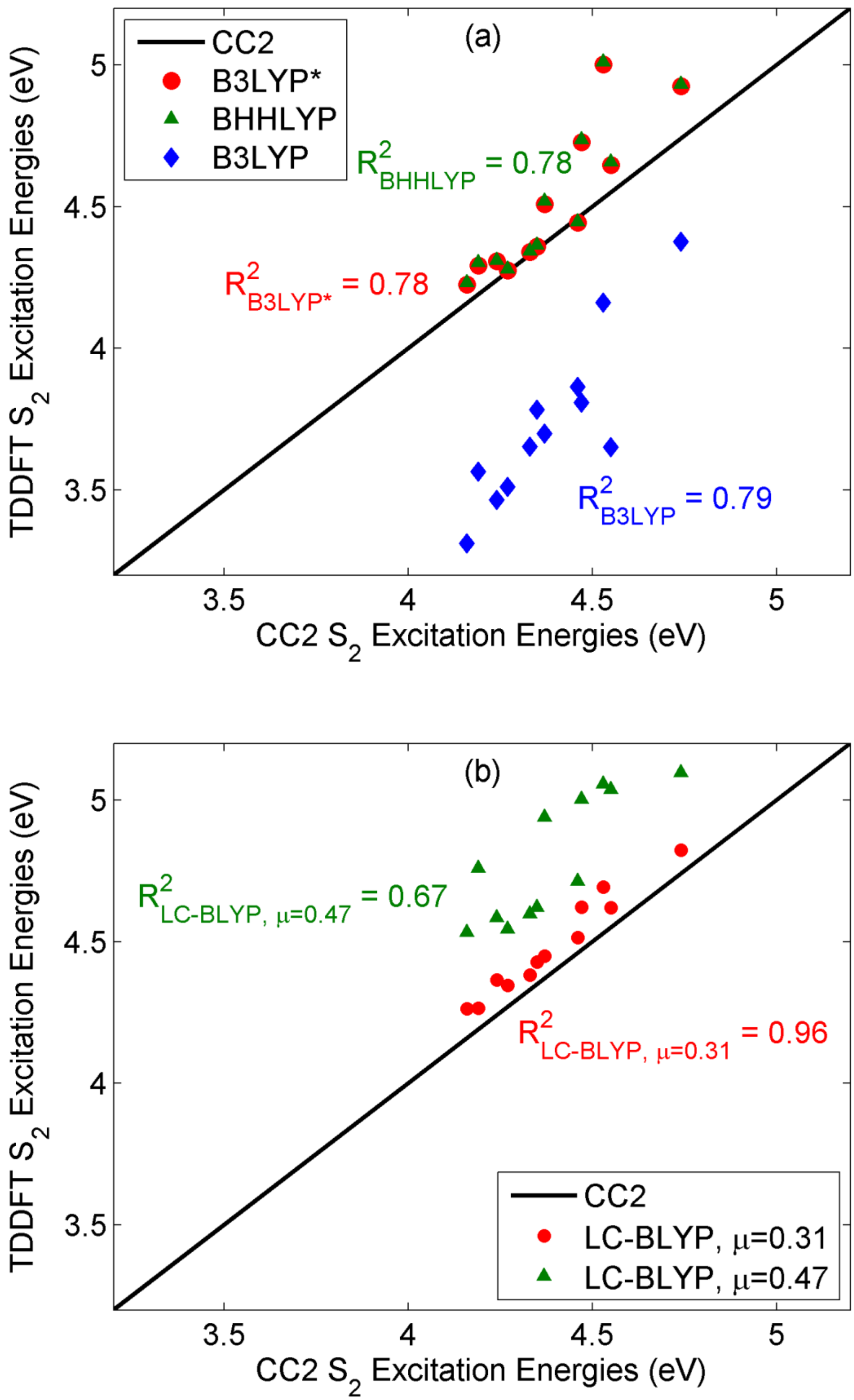

**Fig. 7** Comparison between TDDFT and CC2 $S_2 \leftarrow S_0$ excitation energies for (a) conventional hybrid functionals and (b) range-separated LC-BLYP functionals. The diagonal line in each figure represents a perfect match between CC2 and TDDFT $S_2 \leftarrow S_0$ excitation energies. The $R^2$ values were obtained from a simple linear fit to the data points themselves and not calculated with respect to the diagonal lines shown in the figures. . In Fig. 7(a), the B3LYP* functional gives nearly identical results to the BHHLYP calculations.

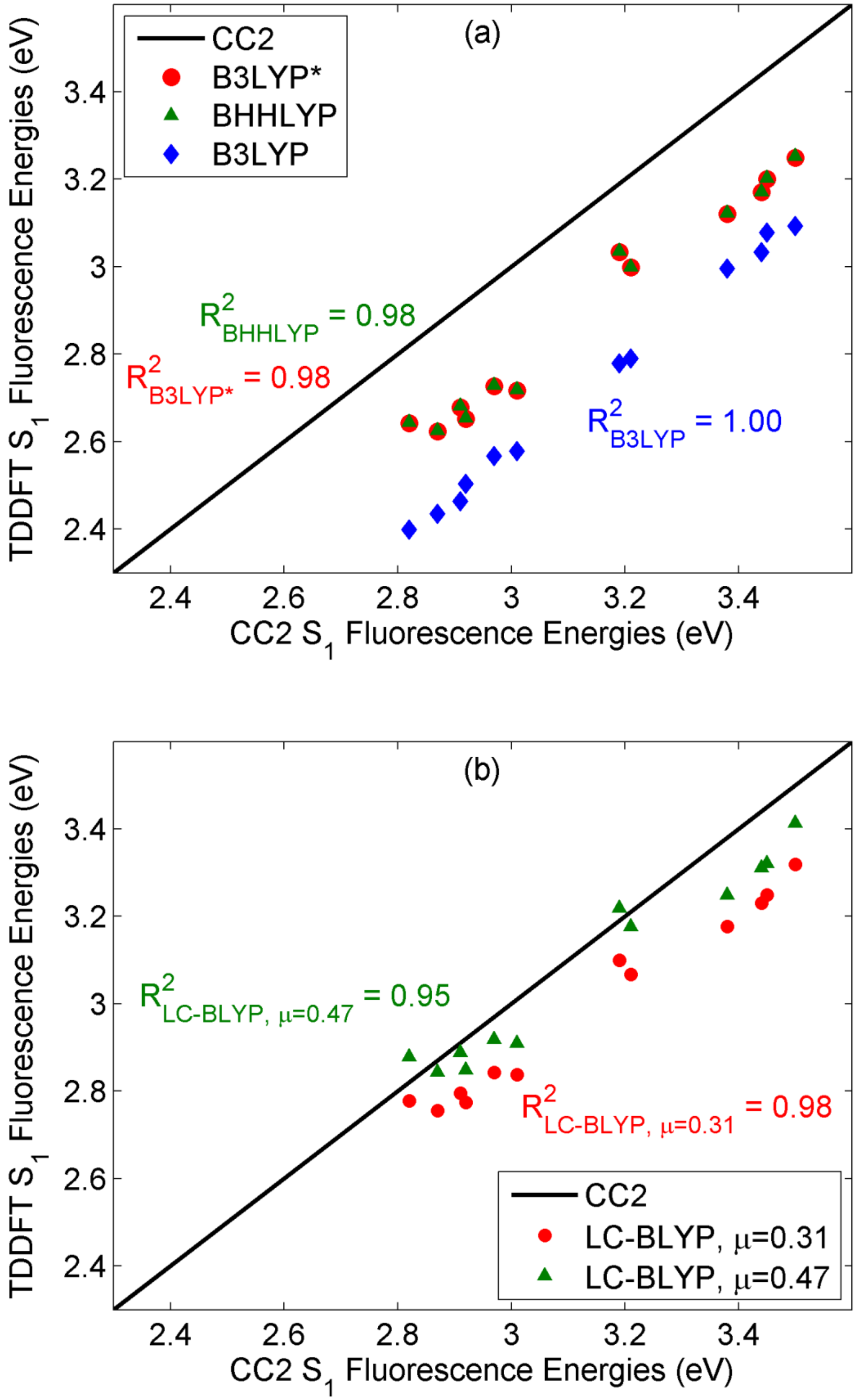

**Fig. 8** Comparison between TDDFT and CC2 $S_1 \rightarrow S_0$ fluorescence energies for (a) conventional hybrid functionals and (b) range-separated LC-BLYP functionals. The diagonal line in each figure represents a perfect match between CC2 and TDDFT $S_1 \rightarrow S_0$ fluorescence energies. The $R^2$ values were obtained from a simple linear fit to the data points themselves and not calculated with respect to the diagonal lines shown in the figures. In Fig. 8(a), the B3LYP* functional gives nearly identical results to the BHHLYP calculations.

**Table 1** $S_1/S_2 \leftarrow S_0$ excitation energies and oscillator strengths for the bithiophene systems. All properties were computed with the ATZVP basis on B3LYP/TZVP-optimized geometries. The B3LYP* functional denotes B3LYP with the RMSE-optimized parameters of $a_0 = 0.49$ and $a_x = 1 - a_0 = 0.51$, which are discussed in section 3.1.

| | | B3LYP ($a_0 = 0.20$) | | B3LYP* ($a_0 = 0.49$) | | LC-BLYP ($\mu = 0.31$) | | CC2 | |
|---|---|---|---|---|---|---|---|---|---|
| System | State | $E_{abs}$ (eV) | Osc. strength | $E_{abs}$ (eV) | Osc. strength | $E_{abs}$ (eV) | Osc. strength | $E_{abs}$ (eV)[a] | Osc. strength[a] |
| NS-[2T] | $S_1$ | **3.54** | 0.61 | **3.82** | 0.64 | **3.90** | 0.63 | **4.01** | 0.70 |
| | $S_2$ | **4.38** | 0.00 | **4.92** | 0.02 | **4.82** | 0.03 | **4.74** | 0.01 |
| NS-[2T]-$S_\alpha$ | $S_1$ | **3.38** | 0.59 | **3.72** | 0.79 | **3.81** | 0.77 | **3.90** | 0.85 |
| | $S_2$ | **3.65** | 0.16 | **4.65** | 0.01 | **4.62** | 0.01 | **4.55** | 0.01 |
| NS-[2T]-$S_\beta$ | $S_1$ | **3.23** | 0.10 | **3.72** | 0.53 | **3.80** | 0.51 | **3.90** | 0.54 |
| | $S_2$ | **3.57** | 0.45 | **4.29** | 0.06 | **4.27** | 0.07 | **4.19** | 0.11 |
| BC-[2T] | $S_1$ | **3.68** | 0.51 | **3.95** | 0.53 | **4.04** | 0.50 | **4.15** | 0.55 |
| | $S_2$ | **4.16** | 0.02 | **5.00** | 0.01 | **4.69** | 0.02 | **4.53** | 0.05 |
| BC-[2T]-$S_\alpha$ | $S_1$ | **3.53** | 0.58 | **3.83** | 0.67 | **3.92** | 0.64 | **4.03** | 0.72 |
| | $S_2$ | **3.81** | 0.06 | **4.73** | 0.00 | **4.62** | 0.01 | **4.47** | 0.03 |
| BC-[2T]-$S_\beta$ | $S_1$ | **3.47** | 0.20 | **3.83** | 0.45 | **3.91** | 0.43 | **4.01** | 0.45 |
| | $S_2$ | **3.70** | 0.25 | **4.51** | 0.02 | **4.45** | 0.03 | **4.37** | 0.08 |

[a]Excitation energies and oscillator strengths from Ref. 47

**Table 2** $S_1/S_2 \leftarrow S_0$ excitation energies and oscillator strengths for the terthiophene systems. All properties were computed with the ATZVP basis on B3LYP/TZVP-optimized geometries. The B3LYP* functional denotes B3LYP with the RMSE-optimized parameters of $a_0$ = 0.49 and $a_x = 1 - a_0$ = 0.51, which are discussed in section 3.1.

| | | B3LYP ($a_0$ = 0.20) | | B3LYP* ($a_0$ = 0.49) | | LC-BLYP ($\mu$ = 0.31) | | CC2 | |
|---|---|---|---|---|---|---|---|---|---|
| System | State | $E_{abs}$ (eV) | Osc. strength | $E_{abs}$ (eV) | Osc. strength | $E_{abs}$ (eV) | Osc. strength | $E_{abs}$ (eV)[a] | Osc. strength[a] |
| NS-[3T] | $S_1$ | **2.94** | 0.93 | **3.25** | 1.02 | **3.41** | 0.99 | **3.45** | 1.11 |
| | $S_2$ | **3.78** | 0.07 | **4.36** | 0.00 | **4.43** | 0.00 | **4.35** | 0.02 |
| NS-[3T]-$S_\alpha$ | $S_1$ | **2.90** | 1.05 | **3.22** | 1.17 | **3.38** | 1.14 | **3.42** | 1.28 |
| | $S_2$ | **3.51** | 0.03 | **4.28** | 0.01 | **4.35** | 0.01 | **4.27** | 0.02 |
| NS-[3T]-$S_\beta$ | $S_1$ | **2.90** | 0.86 | **3.21** | 0.98 | **3.36** | 0.96 | **3.40** | 1.09 |
| | $S_2$ | **3.31** | 0.07 | **4.22** | 0.02 | **4.26** | 0.02 | **4.16** | 0.01 |
| BC-[3T] | $S_1$ | **3.07** | 0.90 | **3.36** | 0.94 | **3.51** | 0.90 | **3.59** | 1.05 |
| | $S_2$ | **3.86** | 0.01 | **4.44** | 0.00 | **4.51** | 0.00 | **4.46** | 0.00 |
| BC-[3T]-$S_\alpha$ | $S_1$ | **2.99** | 1.03 | **3.28** | 1.07 | **3.43** | 1.02 | **3.49** | 1.20 |
| | $S_2$ | **3.65** | 0.01 | **4.34** | 0.01 | **4.38** | 0.00 | **4.33** | 0.01 |
| BC-[3T]-$S_\beta$ | $S_1$ | **2.97** | 0.81 | **3.25** | 0.85 | **3.40** | 0.81 | **3.49** | 0.93 |
| | $S_2$ | **3.46** | 0.00 | **4.31** | 0.01 | **4.37** | 0.01 | **4.24** | 0.01 |

[a]Excitation energies and oscillator strengths from Ref. 47

**Table 3** $S_1 \rightarrow S_0$ fluorescence energies and oscillator strengths for all 12 oligothiophene biomarkers. All properties were computed with the ATZVP basis on TDDFT B3LYP/TZVP-optimized geometries of the $S_1$ state. The B3LYP* functional denotes B3LYP with the RMSE-optimized parameters of $a_0 = 0.49$ and $a_x = 1 - a_0 = 0.51$, which are discussed in section 3.1.

| | B3LYP ($a_0 = 0.20$) | | B3LYP* ($a_0 = 0.49$) | | LC-BLYP ($\mu = 0.31$) | | CC2 | |
|---|---|---|---|---|---|---|---|---|
| System | $E_{fl}$ (eV) | Osc. strength | $E_{fl}$ (eV) | Osc. strength | $E_{fl}$ (eV) | Osc. strength | $E_{fl}$ (eV)[a] | Osc. strength[a] |
| NS-[2T] | **3.08** | 0.64 | **3.20** | 0.64 | **3.25** | 0.61 | **3.45** | 0.73 |
| NS-[2T]-$S_\alpha$ | **2.78** | 0.68 | **3.03** | 0.76 | **3.10** | 0.73 | **3.19** | 0.81 |
| NS-[2T]-$S_\beta$ | **3.00** | 0.55 | **3.12** | 0.57 | **3.18** | 0.55 | **3.38** | 0.67 |
| BC-[2T] | **3.03** | 0.51 | **3.17** | 0.53 | **3.23** | 0.50 | **3.44** | 0.57 |
| BC-[2T]-$S_\alpha$ | **2.79** | 0.63 | **3.00** | 0.67 | **3.07** | 0.64 | **3.21** | 0.75 |
| BC-[2T]-$S_\beta$ | **3.09** | 0.48 | **3.25** | 0.50 | **3.32** | 0.48 | **3.50** | 0.56 |
| NS-[3T] | **2.57** | 1.02 | **2.73** | 1.05 | **2.84** | 1.00 | **2.97** | 1.18 |
| NS-[3T]-$S_\alpha$ | **2.40** | 1.09 | **2.64** | 1.21 | **2.78** | 1.15 | **2.82** | 1.31 |
| NS-[3T]-$S_\beta$ | **2.46** | 0.87 | **2.68** | 0.99 | **2.80** | 0.94 | **2.91** | 1.02 |
| BC-[3T] | **2.58** | 0.96 | **2.72** | 0.97 | **2.84** | 0.91 | **3.01** | 1.11 |
| BC-[3T]-$S_\alpha$ | **2.43** | 1.08 | **2.62** | 1.12 | **2.76** | 1.06 | **2.87** | 1.27 |
| BC-[3T]-$S_\beta$ | **2.50** | 0.85 | **2.65** | 0.88 | **2.77** | 0.83 | **2.92** | 1.03 |

[a]Fluorescence energies from Ref. 47

**Table 4** Ground- and excited-state dipole moments associated with the $S_1/S_2 \leftarrow S_0$ absorption transitions. All dipoles were computed with the ATZVP basis on B3LYP/TZVP-optimized geometries. The B3LYP* functional denotes B3LYP with the RMSE-optimized parameters of $a_0 = 0.49$ and $a_x = 1 - a_0 = 0.51$, which are discussed in section 3.1.

| | B3LYP ($a_0 = 0.20$) | | | B3LYP* ($a_0 = 0.49$) | | | LC-BLYP ($\mu = 0.31$) | | |
|---|---|---|---|---|---|---|---|---|---|
| System | $S_0$ dipole (D) | $S_1$ dipole (D) | $S_2$ dipole (D) | $S_0$ dipole (D) | $S_1$ dipole (D) | $S_2$ dipole (D) | $S_0$ dipole (D) | $S_1$ dipole (D) | $S_2$ dipole (D) |
| NS-[2T] | 1.79 | 5.78 | 7.05 | 1.67 | 4.47 | 1.90 | 1.45 | 4.27 | 1.74 |
| NS-[2T]-$S_\alpha$ | 1.91 | 8.53 | 13.79 | 1.89 | 4.32 | 11.64 | 1.77 | 3.99 | 8.55 |
| NS-[2T]-$S_\beta$ | 2.53 | 9.31 | 6.08 | 2.38 | 4.60 | 8.39 | 2.21 | 4.42 | 7.45 |
| BC-[2T] | 4.20 | 5.20 | 7.04 | 4.36 | 5.22 | 3.89 | 4.21 | 4.87 | 3.75 |
| BC-[2T]-$S_\alpha$ | 3.33 | 5.43 | 11.78 | 3.44 | 4.13 | 9.69 | 3.35 | 3.81 | 2.87 |
| BC-[2T]-$S_\beta$ | 3.57 | 6.96 | 7.19 | 3.69 | 4.33 | 8.90 | 3.54 | 4.02 | 8.16 |
| NS-[3T] | 2.25 | 8.91 | 2.57 | 1.95 | 5.99 | 3.35 | 1.56 | 5.09 | 3.17 |
| NS-[3T]-$S_\alpha$ | 1.62 | 8.47 | 22.31 | 1.40 | 4.91 | 3.21 | 1.26 | 3.93 | 3.12 |
| NS-[3T]-$S_\beta$ | 2.88 | 8.37 | 14.35 | 2.59 | 5.64 | 6.30 | 2.33 | 4.85 | 5.85 |
| BC-[3T] | 3.62 | 6.31 | 3.50 | 3.73 | 5.33 | 4.07 | 3.54 | 4.72 | 3.83 |
| BC-[3T]-$S_\alpha$ | 2.67 | 4.49 | 17.34 | 2.73 | 3.70 | 3.30 | 2.63 | 3.28 | 2.73 |
| BC-[3T]-$S_\beta$ | 5.25 | 5.81 | 12.48 | 5.39 | 5.81 | 6.13 | 5.28 | 5.52 | 5.31 |

**Table 5** Ground- and excited-state dipole moments associated with the $S_1 \rightarrow S_0$ fluorescence transition. All dipoles were computed with the ATZVP basis on TDDFT B3LYP/TZVP-optimized geometries of the $S_1$ state. The B3LYP* functional denotes B3LYP with the RMSE-optimized parameters of $a_0 = 0.49$ and $a_x = 1 - a_0 = 0.51$, which are discussed in section 3.1.

| | B3LYP ($a_0 = 0.20$) | | B3LYP* ($a_0 = 0.49$) | | LC-BLYP ($\mu = 0.31$) | |
|---|---|---|---|---|---|---|
| System | $S_0$ dipole (D) | $S_1$ dipole (D) | $S_0$ dipole (D) | $S_1$ dipole (D) | $S_0$ dipole (D) | $S_1$ dipole (D) |
| NS-[2T] | 2.43 | 3.88 | 2.32 | 3.65 | 1.97 | 3.84 |
| NS-[2T]-$S_\alpha$ | 2.76 | 7.58 | 2.41 | 5.76 | 1.99 | 5.76 |
| NS-[2T]-$S_\beta$ | 2.92 | 3.76 | 2.80 | 3.66 | 2.51 | 3.73 |
| BC-[2T] | 4.34 | 4.62 | 4.53 | 5.02 | 4.32 | 4.90 |
| BC-[2T]-$S_\alpha$ | 3.48 | 6.37 | 3.52 | 5.44 | 3.19 | 5.34 |
| BC-[2T]-$S_\beta$ | 3.54 | 4.07 | 3.67 | 4.26 | 3.43 | 4.04 |
| NS-[3T] | 3.21 | 6.29 | 2.95 | 5.39 | 2.36 | 5.23 |
| NS-[3T]-$S_\alpha$ | 4.26 | 10.67 | 3.70 | 7.98 | 3.05 | 7.37 |
| NS-[3T]-$S_\beta$ | 4.96 | 9.19 | 4.64 | 7.54 | 4.03 | 7.33 |
| BC-[3T] | 3.98 | 5.08 | 4.14 | 5.11 | 3.83 | 4.85 |
| BC-[3T]-$S_\alpha$ | 4.95 | 7.85 | 4.95 | 6.75 | 4.65 | 6.33 |
| BC-[3T]-$S_\beta$ | 5.23 | 5.66 | 5.39 | 5.80 | 5.20 | 5.61 |